\documentstyle[prd,psfig,aps,amsfonts,twocolumn]{revtex}

\catcode`\@=11

\def\maketitle2{\par 
\begingroup
\let\cite\@bylinecite
\def\thefootnote{\fnsymbol{footnote}}%
\twocolumn[\@maketitle2\vskip2pc]%
\thispagestyle{plain}\@thanks
\endgroup
\def\thefootnote{\arabic{footnote}}%
\setcounter{footnote}{0}%
\let\maketitle2\relax \let\@maketitle2\relax
\let\@thanks\relax \let\@authoraddress\relax \let\@title\relax
\let\@date\relax \let\thanks\relax \let\@abstract\relax 
\let\@pacs\relax}

\def\abstract#1{\gdef\@abstract{{\par 
\bgroup
\ifdim\prevdepth=-1000pt \prevdepth0pt\fi
\hsize\columnwidth
\dimen0=-\prevdepth \advance\dimen0 by17.5pt \nointerlineskip
\small\vrule width 0pt height\dimen0 \relax}{~~}#1\egroup}}

\def\pacs#1{\gdef\@pacs{{\par 
\bgroup
\hsize\columnwidth \parindent0pt
\ifdim\prevdepth=-1000pt \prevdepth0pt\fi
\dimen0=-\prevdepth \advance\dimen0 by20pt\nointerlineskip
\egroup} PACS numbers:~#1}}

\def\@maketitle2{
\@preprint
\@title
\ifdim\prevdepth=-1000pt \prevdepth0pt\fi
\@authoraddress
\@date
\begin{list}{}{\leftmargin=0.10753\textwidth \rightmargin=\leftmargin
\itemsep=1pc\partopsep=-1pc}
\item\@abstract
\item\@pacs
\end{list}
}

\catcode`\@=12

\newif\ifinlinefigures
\inlinefigurestrue

\newcommand{\sm}[1]{{\scriptscriptstyle #1}}
\def\d{\partial}
\def\l{\left(}
\def\r{\right)}

\newcommand{\Tr}{\mathop{\rm Tr}\nolimits}
\renewcommand{\Im}{\mathop{\rm Im}\nolimits}

\newcommand{\be}{\begin{equation}}
\newcommand{\ee}{\end{equation}}
\newcommand{\ba}{\begin{eqnarray}}
\newcommand{\ea}{\end{eqnarray}}
\begin{document}
\draft
\flushbottom
\title{Induced charge matching and  Wess--Zumino term on
  quantum modified moduli space}
\author{S.~L.~Dubovsky, D.~S.~Gorbunov }
\address{Institute for Nuclear Research of
         the Russian Academy of Sciences,  60th October Anniversary
  Prospect, 7a, 117312 Moscow, 
Russia}
\abstract{
  Recently it was proposed that matching of global charges induced in vacuum
 by 
 slowly varying, topologically non-trivial scalar fields provides 
 consistency conditions analogous to the 't Hooft anomaly matching
 conditions. We study matching of induced charges in supersymmetric
 $SU(N)$ gauge theories with quantum modified moduli space. We find that
 the Wess--Zumino term should be present in the low energy theory
  in order that these
 consistency conditions are satisfied. 
  We calculate the lowest homotopy groups of the
 quantum moduli space, and show that there are 
 no obstructions to the existence of the Wess--Zumino
 term at arbitrary $N$.  The explicit expression
 for this term is given. It is shown that neither vortices nor
 topological solitons
 exist in the model. The case of softly broken
 supersymmetry is considered as well. 
 We find that the possibility of the global flavor symmetric
 vacuum is strongly disfavored.}
\maketitle2
\narrowtext
\section{Introduction}
\label{intro}
Low-energy dynamics of gauge theories strongly coupled in the
infrared often can be  described in terms of effective theories. For
instance, the soft limit of quantum chromodynamics (QCD) 
is the theory of self-interacting pion
fields. In more sophisticated models (e.g., supersymmetric gauge
theories), the effective low-energy theories exhibit interesting
content of composite fermions as well. This property may be important for
the construction of composite models of quarks and leptons.

Most of the results concerning the infrared behavior
of supersymmetric gauge theories are obtained without detailed
calculations which are problematic in the strong coupling
domain. Instead, one makes use of general properties, such as
symmetries and holomorphy. 
One of the most powerful tools for constraining the
low-energy spectra is provided by the 't Hooft anomaly matching
conditions~\cite{'thooft}.  Namely, one introduces topologically
non-trivial background gauge fields corresponding to the flavor
symmetry group and checks that the anomalies in global currents are
the same in the microscopic and effective theories. The basis for
the anomaly matching conditions is provided by the Adler-Bardeen
non-renormalization theorem~\cite{adler}.

Recently, it was suggested to consider topologically non-trivial,
slowly varying in space scalar fields as other probes of 
effective theories~\cite{rubakov}. The corresponding
matching conditions emerge due to
the fact that such background 
scalar fields generically induce global 
charges in
vacuum. These charges are the quantities which should match in the
microscopic and low energy descriptions of the model. The 
non-renormalization theorem justifying these matching
conditions was proven in Ref.~\cite{weandru}.

Induced charge matching in supersymmetric QCD (SQCD)
with $SU(N_c)$ gauge group and the number of
flavors $N_f$  larger than the number of colors has been discussed
in Refs.~\cite{rubakov,weandru}. Models with softly broken supersymmetry
were also considered there and it was found that in the latter case 
the induced charge matching
conditions provide new information in addition to the 't Hooft
conditions. The present paper is devoted to the study of 
induced charge matching in SQCD at $N_f=N_c$ when the quantum
deformation of the moduli space takes place~\cite{seiberg}.
It is shown that all constraints are satisfied provided the term
analogous to the Wess--Zumino term in QCD is added. This term has been already
constructed  in the case of $SU(2)$ gauge 
group in Ref.~\cite{manohar}. Here
we extend this construction to the case of $SU(N_c)$ group with
arbitrary $N_c$. In the case of
softly broken supersymmetry it is found that induced charge matching
conditions require that  flavor symmetry group is broken in the
global vacuum.

The paper is organized as follows. In section \ref{matching} 
we recapitulate the notion of
induced charge matching. We consider the QCD case and discuss
the role of the Wess--Zumino term there. In section \ref{SQCD} we study
induced charge matching in SQCD with quantum modified moduli
space, analyze its topology and construct the analogue of the
Wess--Zumino term in this model. The case of softly broken
supersymmetry is considered in section \ref{break}.
In  section \ref{Conclusions} we
present our conclusions.  Appendix A is devoted to the details of the
topological analysis of the quantum moduli space and in Appendix B we
show that 
the Wess--Zumino term of SQCD is unambiguous in quantum theory.
\section{Induced charge matching}
\label{matching}
To begin with, 
let us consider QCD with $N_c$ colors and $N_f$ massless
fermion flavors.  Let $\psi^a$ and $\tilde{\psi}_{\tilde{a}}$,
$a,\tilde{a}=1,\ldots,N_f$, denote left-handed quarks and anti-quarks,
respectively. This theory exhibits the global $SU(N_f)_L\times
SU(N_f)_R$ symmetry and non-anomalous baryon symmetry, $\psi\to{\rm
e}^{i\alpha}\psi$, $\tilde{\psi}\to{\rm e}^{-i\alpha}\tilde{\psi}$.
To probe the theory, one introduces time-independent background scalar fields
$m^{\tilde{a}}_{b}({\bf x})$ of the following form,
\begin{equation}
\label{mass}
m^{\tilde{a}}_{b}({\bf x})=m_0U^{\tilde{a}}_{b}({\bf x})\;,
\end{equation}
where $m_0$ is a constant and $U^{\tilde{a}}_b({\bf x})$ is an $SU(N_f)$
matrix at each point ${\bf x}$.

Let these fields interact with quarks and anti-quarks,
\begin{equation}
{\cal L}_{\rm int}=\tilde{\psi}_{\tilde{a}}m^{\tilde{a}}_b\psi^b+{\rm h.c.}
\label{V*}
\end{equation}
In what follows we restrict the form of the background
fields by requiring that $U({\bf x})$ 
tends to a constant at spatial
infinity. One can always make this constant  equal to unity,
$U({\bf x})\to {\bf 1}$ as $|{\bf x}|\to \infty$,
by making use of a global  $SU(N_f)$ rotation.

The baryonic current is conserved and obtains non-vanishing
vacuum expectation value in the presence of the background scalar fields.
To the leading order in momenta, the one-loop result for the
induced current is~\cite{GW}
\begin{equation}
\label{complete}
\langle j_B^\mu\rangle = \frac{N_c}{24 \pi^2}  \epsilon^{\mu\nu\lambda\rho}
           \Tr \left( U\partial_{\nu} U^{\dagger}
                 U\partial_{\lambda} U^{\dagger}  U\partial_{\rho}
            U^{\dagger}        
            \right)\;.
\end{equation}
It follows from this expression that the baryonic
charge induced in vacuum is proportional to the topological number
of the background,
\begin{equation}
\label{ch}
\langle B\rangle = N_c N[U] \;,
\end{equation}
where 
\begin{equation}
\label{NU}
N[U]=\frac{1}{24 \pi^2} \int\!d^3 x~ \epsilon^{ijk}
           \Tr \left( U\partial_i U^{\dagger}
                 U\partial_j U^{\dagger}  U\partial_k U^{\dagger}        
            \right)\;.
\end{equation}
This topological property is the basis of the theorem proven in
Ref.~\cite{weandru} that states that Eq.~(\ref{ch}) does not get
renormalized in the full quantum theory provided the expansion in
momenta of the background fields is valid. 
It is worth noting also, that
higher derivative terms omitted in Eq.~(\ref{complete}) do not
contribute to $\langle B\rangle$. 

More generally, one does 
not necessarily introduce masses to all quarks, and 
considers instead background fields interacting only with some of
the flavors. The only requirement is that all fermions coupled to
the background fields 
become massive due to this interaction. Instead
of the baryon number one may study generators of other
non-anomalous global
symmetries which 
remain unbroken in the presence of the background fields.

The non-renormalization theorem implies that the
induced charges should match
in the microscopic and  effective theories.
These matching conditions are analogous, but generally inequivalent,
to the 't Hooft conditions.

Let us see how induced charges match in QCD with $N_f\ge 3$.
Following Ref.~\cite{rubakov} let us consider a general case when a
mass term (\ref{V*}) is added to the first $N_0$ quark flavors.
There are two independent generators of the original $SU(N_f)_L\times
SU(N_f)_R\times U(1)_B$ symmetry group that remain unbroken in the
presence of the background fields and act non-trivially on the first
 $N_0$ quark flavors. The first one is a generator of a baryon
symmetry and the second one 
transforms first $N_0$ flavors in the same way as
baryon symmetry and acts trivially on the other quarks.

As all fundamental fermions that couple to the background scalar
field acquire masses due to this interaction, the derivative expansion
is justified. Hence, for slowly varying $m({\bf x})$ one has
\begin{equation}
\label{QH}
\langle Q_i \rangle= N_c N[U]
\end{equation}
for the induced charges $\langle Q_i \rangle$ corresponding to the
above  two
 generators.

The low energy dynamics of QCD is described in terms of 
$SU(N_f)$-valued sigma-model field $V(x)$. The interaction 
with the background field $m({\bf x})$ induces a potential term
into the low-energy effective Lagrangian,
\[
           \Delta {\cal L}_{\rm eff} =
         \Tr\left( m^{\dagger}V + V^{\dagger}m \right)\;.
\]
For slowly varying $m({\bf x})$, 
the effective potential is  minimized at
\begin{equation}
 V({\bf x}) =
\left(
\begin{array}{cc}
   U({\bf x})  & 0 \\
   0 &  {\bf 1}
\end{array}
\right)\;.   
\label{6*}
\end{equation}
As we will see shortly, the induced charge matching conditions 
are satisfied provided the  Wess--Zumino term \cite{wess} is added 
in the effective action.
This term
cannot be written as a four-dimensional integral of a
$SU(N_f)_L\times SU(N_f)_R$-invariant non-singular function and is
defined as follows~\cite{witten1}. A field configuration $V(x)$ of
finite energy defines a map $S^4\to SU(N_f)$ from compactified
spacetime to the space of vacua (moduli space). Since
$\pi_4(SU(N_f))=0$, the image of this map is a boundary of a
five-dimensional submanifold $\Sigma_5$ in $SU(N_f)$.  The Wess--Zumino
term is the integral over this surface,
\begin{eqnarray}
\label{wesszum}
&&\Gamma_{QCD}={-iN_c\over 240\pi^2}\int_{\Sigma_5}\!\!\! d\Omega~
\epsilon^{\mu\nu\lambda\rho\sigma}\\&&
\times\Tr\l V^{-1}\partial_{\mu}VV^{-1}\partial_{\nu}VV^{-1}\partial_{\lambda}V
V^{-1}\partial_{\rho}VV^{-1}\partial_{\sigma}V\r\;.\nonumber
\end{eqnarray}
This expression is invariant under small deformations of $\Sigma_5$
because the integrand is a closed five-form.  The discrete ambiguity
in the definition of $\Gamma_{QCD}$ is related to non-zero homotopy
group $\pi_5(SU(N_f))$. However, this ambiguity is irrelevant in
quantum theory, because $e^{i\Gamma_{QCD}}$ is well-defined.  The
contribution of the Wess--Zumino term to all global currents except 
for
baryonic one can be
obtained by making use of 
the conventional Noether procedure and reads
\cite{witten1}
\begin{eqnarray}
\label{wzcurrent}
j^{\mu a}={1\over 48\pi^2}\epsilon^{\mu\nu\lambda\rho}\Tr\l T_L^a
V^{-1}\partial_{\nu}VV^{-1}\partial_{\lambda}V
V^{-1}\partial_{\rho}V\right.\nonumber\\\left.
+T_R^a
\partial_{\nu}VV^{-1}\partial_{\lambda}V
V^{-1}\partial_{\rho}VV^{-1}\r\;,
\end{eqnarray}
where $T_L^a$ and $T_R^a$ are  $SU(N_f)_L$ and $SU(N_f)_R$ components of a
generator of a symmetry transformation.  Comparing Eq.~(\ref{wzcurrent})
with the formulae (\ref{QH}) and (\ref{NU}) for the induced charges
and taking into account the vacuum value (\ref{6*}) of the field
$V(x)$, one checks that induced charges match for
the unbroken generator of the flavor group
$SU(N_f)_L\times SU(N_f)_R$. The expression for the induced
baryonic current can be obtained by the formal substitution $T^a_{L,R}\to
{\bf 1}$ in  Eq.~(\ref{wzcurrent}), see Ref.~\cite{witten1}
for details. 
Again, due to Eq.~(\ref{6*}) induced baryonic charges match in the
fundamental and effective theories.
We conclude that induced charges match in QCD
due to the presence of the Wess--Zumino term.

\section{Wess--Zumino term in SQCD}
\label{SQCD}
Let us turn now to the case of SQCD with $N_f=N_c=N$. This
theory exhibits quantum deformation
of the moduli space~\cite{seiberg}. Namely, the space of vacua of the
microscopic theory is described by the set of holomorphic gauge invariants
constructed out 
of quarks $Q^a$ and anti-quarks $\tilde{Q}_{\tilde{b}}$. These
invariants are mesons,
\[
M^a_{\tilde{b}}=Q^a\tilde{Q}_{\tilde{b}}
\]
and (anti)-baryons
\[
B=\epsilon_{a_1..a_{{\sm{N}}}}Q^{a_1}..Q^{a_{{\sm{N}}}}\;,
\]
\[
\tilde{B}=\epsilon^{\tilde{b}_1..\tilde{b}_
{{\sm{N}}}}\tilde{Q}_{\tilde{b}_1}..\tilde{Q}_{\tilde{b}_{{\sm{N}}}} 
\]
subject to the constraint
\be
\label{qms}
\det\! M-B\tilde{B}=\Lambda^{2N}\;,
\ee 
where $\Lambda$ is the infrared scale of the theory.
The r.h.s of 
Eq.~(\ref{qms}) is of purely quantum origin and indicates the difference 
between the topologies of the quantum and classical spaces of vacua.

In the low-energy theory of mesons and (anti-)baryons the constraint
(\ref{qms}) can be presented as  an effective superpotential
\be
\label{potqms}
W={\cal A}(\det\! M-B\tilde{B}-\Lambda^{2N})\;,
\ee
where ${\cal A}$ is  the Lagrange multiplier superfield.

Let us probe this theory by adding the scalar background field 
$m^{\tilde{q}}_p({\bf x})$ with the same properties as above, i.e.,
by introducing the term
\begin{equation} 
   m^{\tilde{b}}_a({\bf x}) \tilde{Q}_{\tilde{b}} Q^a    
\label{31*} 
\end{equation}
into the superpotential of the fundamental theory. We add
the mass terms to {\it all} quark flavors to
avoid the run-away vacuum. Then for the general matrix
$U^{\tilde{b}}_a({\bf x})$, the external fields are neutral only under
the baryon symmetry\footnote{In section~\ref{break} we discuss a special
case where additional  unbroken global 
symmetries are present.}. The
calculation of the induced baryonic charge in the fundamental theory
proceeds as in sect. 2, and we again obtain
\be
\label{bf}
\langle B\rangle=N_cN[U]\;.
\ee

We now turn to effective low energy theory.
For slowly varying $m({\bf x})$ the interaction (\ref{31*}) translates
into the additional term $\Tr mM$ in the effective
superpotential (\ref{potqms}). As a result, the ground state is
described by the following ${\bf x}$-dependent expectation
values\footnote{Hereafter we use the same notations for superfields
and their scalar components.} of mesons and baryons,
\ba
\label{vevs}
M({\bf x})=\Lambda^2U^{-1}({\bf x})\;,
\\ B=\tilde{B}=0\nonumber\;.  
\ea
One can explicitly check that no baryonic
charge is induced due to
superpotential interaction in the effective theory, as opposed to
the
supersymmetric models considered in Refs.~\cite{rubakov,weandru}. So,
the Wess--Zumino term is needed
in complete analogy to the QCD case.

The Wess--Zumino term has been already 
constructed in the case of $SU(2)$
group~\cite{manohar}. In that theory one combines 
mesons and baryons into a single $4\times 4$ anti-symmetric matrix,
\begin{equation}
V=\left(\begin{array}{c|c}
\begin{array}{cc}
0 & B \\
-B & 0 \\
\end{array} & M \\ \hline
- M^T & \begin{array}{cc}
0 & \tilde B \\
-\tilde B & 0 \\
\end{array}
\end{array}\right)\;.
\end{equation}
 In terms of this matrix, the Wess--Zumino term
can be written as follows\footnote{Note that the matrix $V$
is non-degenerate due to the constraint (\ref{qms}).},
\ba
\label{manoh}
&&\Gamma_{SU(2)}={1\over 240\pi^2}\Im\int_{\Sigma_5}\!\!\! d\Omega~
\epsilon^{\mu\nu\lambda\rho\sigma}\\&&
\times\Tr\l V^{-1}\partial_{\mu}VV^{-1}\partial_{\nu}VV^{-1}\partial_{\lambda}V
V^{-1}\partial_{\rho}VV^{-1}\partial_{\sigma}V\r\;.\nonumber
\ea
In fact, this equation defines only the bosonic part of the Wess--Zumino
action. However, it is sufficient for our purposes, since it is this
part that is relevant for charge matching. 
The Wess--Zumino term (\ref{manoh}) is written in the holomorphic form,
see Ref.~\cite{holomor} for
a discussion of why this is possible.

As pointed out in Ref.~\cite{manohar}, the generalization of 
Eq.~(\ref{manoh}) to the case of $SP(N)$ group is straightforward. 
So, let us consider $SU(N)$ gauge group at $N \geq 3$.

In
order that the Wess--Zumino term  could be constructed, 
the fourth homotopy group of the quantum
moduli space
$\cal Q$ determined by Eq.~(\ref{qms})
should be trivial, $\pi_4({\cal Q})=0$.
Moreover, if $\pi_5({\cal Q})$ is non-trivial, the value of the Wess--Zumino
functional $\Gamma$ on its generators should be equal to $2\pi n$, $n\in
\mathbb{Z}$, so that $e^{i\Gamma}$ is well-defined in quantum
theory.

In Appendix A it
is shown  that 
\be
\label{susp}
{\cal Q}\sim \Sigma\l\Sigma\l SU(N)\r\r\;.  
\ee 
Here by ``$\sim$'' we
denote homotopic equivalence and $\Sigma ({\cal X})$ is a
suspension of the manifold ${\cal X}$ (see, e.g., Ref.~\cite{fux}
for definitions and notations). The
generalized Freidental theorem (\cite{fux}, p.79) states that 
$\pi_{q+1}\l\Sigma\l{\cal
X}\r\r=\pi_{q}\l{\cal X}\r$ for $q\leq 2n-2$, provided $\pi_i({\cal X})=0$
for $i<n$. In particular, Eq.~(\ref{susp}) implies
that $\pi_{1}({\cal Q})=\pi_{2}({\cal Q})=\pi_{3}({\cal Q})=
\pi_{4}({\cal Q})=0$ 
 and $\pi_{5}({\cal
Q})= \mathbb Z$.

Therefore, in the case of $SU(N_c)$ SQCD with $N_f=N_c$, the
Wess--Zumino term can be constructed as a five-dimensional integral,
in similarity
to the QCD case.  Another outcome of our calculation is
that the triviality of the groups
$\pi_2({\cal Q})$ and $\pi_3({\cal Q})$  implies that neither
vortices nor topological solitons exist in this theory.

The explicit expression
generalizing Eq.~(\ref{manoh}) is 
\ba
\label{our}
\Gamma_{SQCD}&=&{-1\over 12\pi^2\Lambda^{4N}}\Im\int_{\Sigma_5}\!\!\! d\Omega~
\det\!M\cdot\epsilon^{\mu\nu\lambda\rho\sigma}
\d_{\mu}B\;\;\!\!\d_{\nu}
\tilde{B}\nonumber\\&&\times \Tr\l
M^{-1}\partial_{\lambda}M
M^{-1}\partial_{\rho}MM^{-1}\partial_{\sigma}M\r\;.  
\ea 
It is straightforward to check that Eq.~(\ref{our}) indeed reproduces
Eq.~(\ref{manoh}) at $N=2$.

It is shown in the Appendix B that $\Gamma_{SQCD}=2\pi$
when $\Sigma_5$ is a generator of $\pi_5({\cal Q})$
and consequently the Wess--Zumino term~(\ref{our})  is
unambiguous  in quantum theory. At first sight
$\Gamma_{SQCD}$
appears singular at the points where $\det\! M=0$. However, this is not
the case. Indeed, due to the 
constraint (\ref{qms}) one has $B,\tilde{B}\neq 0$ at these
points, and the integrand in the r.h.s. of Eq.~(\ref{our}) can be
rewritten in the following way,
\ba
{-1\over 2}\epsilon^{\mu\nu\lambda\rho\sigma}\d_{\mu}\bigg(
{\d_{\nu}\tilde{B}\over\tilde{B}}
\Tr\l\d_{\lambda}MM^{-1}\det\! M\;\!\nonumber\right.\\\times\left.
\d_{\rho}M\d_{\sigma}\l 
M^{-1}\det\! M\r\r\bigg)\;.\nonumber
\ea
This expression is explicitly non-singular when $\det\! M=0$
because $M^{-1}\det\! M$ can be 
defined there by making use of the identity
$M^{-1}\det\! M=\tilde{M}^T$,
where $\tilde{M}$ is a matrix composed of the minors of the matrix $M$.

Let us check that the Wess--Zumino term (\ref{our}) properly
reproduces the induced baryonic charge. Contrary to the QCD case, one
can straightforwardly obtain the contribution of the Wess--Zumino term
to the baryonic current by making use of 
the conventional Noether procedure,
\ba
\label{ourbar}
j^{\mu}_B&=&{N\over 48\pi^2\Lambda^{4N}}\det\! M^2\cdot
\epsilon^{\mu\nu\lambda\rho}\nonumber\\&&
\times\Tr\l\d_{\nu}MM^{-1}\d_{\lambda}MM^{-1}\d_{\rho}MM^{-1}\r+h.c.
\ea
Then, substituting the vacuum expectation value of the meson fields
(\ref{vevs}) one obtains the same value of the induced baryonic charge
as in the microscopic theory, eq. (\ref{bf}). 

To summarize, the above analysis shows that
Eq.~(\ref{our}) is a well-defined expression for the 
Wess--Zumino term in the $SU(N)$ SQCD with quantum modified moduli space and
that induced charge matching conditions are satisfied in this theory when
the term (\ref{our}) is taken into account.

\section{Softly broken SQCD}
\label{break}
Finally, let us discuss induced charge matching in the model with
the soft supersymmetry
breaking mass term
\be
\label{soft}
V_{soft}=\mu_{Q}^2 (|Q|^2+|\tilde{Q}|^2) 
\ee 
added to the potential
of the microscopic theory. It has been suggested in
Ref.~\cite{peskin} that, at least at $\mu_Q\ll\Lambda$, the effective
theory is described both by the constraint~(\ref{qms}) and by the the
soft terms 
\be
\label{softlow}
V_{eff}={\mu_B^2\over
\Lambda^{2N-2}}(|B|^2+|\tilde{B}|^2)+{\mu_M^2\over \Lambda^2}|M|^2 \;.  
\ee 
Then, up to
flavor rotations, there
are two candidates for  global minima of this potential. In the first one
the baryon symmetry is unbroken, $B=\tilde{B}=0$ ,
$M^a_{\tilde{b}}=\Lambda^{2}\cdot\mathbf{1}$, while in the second one the
flavor symmetries are unbroken, $B=-\tilde{B}=\Lambda^{N}$ ,
$M^a_{\tilde{b}}=0$. Which of these points is the global vacuum depends on the
ratio ${\mu_B^2\over \mu_{M}^2}$ that has not been calculated.
Indeed, 
the baryon number violating
stationary point is always a stable minimum, while the stationary
point with $B=\tilde{B}=0$ is unstable if ${\mu_B^2\over \mu_{M}^2}<1$
and  is a global vacuum if ${\mu_B^2\over \mu_{M}^2}>{N\over
2}$. 

Let us now introduce the space-dependent mass term (\ref{31*}). 
The induced baryonic charge as calculated in the microscopic theory 
is the same 
as in the supersymmetric case. In the low-energy theory there are
again two candidates for the global minimum of the effective potential.
The first one corresponds to the stationary point 
with unbroken baryon symmetry
and is the same as in the supersymmetric case (see Eq.~(\ref{vevs})).
In the second  stationary point the baryon symmetry is broken, 
$B=-\tilde{B}=\Lambda^{N}$ , $M^a_{\tilde{b}}=0$. 
The baryonic current induced in the first candidate vacuum 
is given by 
Eq.~(\ref{ourbar}) and properly reproduces the induced charge as
calculated in the
microscopic theory. In the second extremum the baryon symmetry is
spontaneously
broken and  the baryon charge matching
condition cannot be straightforwardly applied. 
Consequently, the matching condition for the baryonic charge
does not enable one 
to  conclude which of the two extrema is the global vacuum.

However, for some special choices of the background scalar field
(\ref{mass}), additional matching conditions arise. Namely, let us take
the matrix $U^a_{\tilde{b}}(\bf{x})$ in the following form,
\[
U(\bf{x})=\left(
\begin{array}{cc}
\tilde{U}(\bf{x}) & 0 \\
0 &  \mathbf 1 \\
\end{array}\right)\;,
\]
where $\tilde{U}(\bf{x})$ is $N_0\times N_0$ unitary matrix. In analogy
to the QCD case, background fields of this form respect
two independent symmetries acting non-trivially on the first
$N_0$ flavors. The first one is the baryon symmetry.  The
second one is a vectorial subgroup of the original $SU(N_f)_L\times
SU(N_f)_R$ flavor group; its generator is
\[
    T^f = \mbox{diag}\left( 1, \dots, 1, -\frac{N_0}{N_f - N_0},\dots,
                  -\frac{N_0}{N_f - N_0}\right)\;.
\]
In the microscopic theory 
both induced charges are equal to $N_cN[\tilde{U}]$.

In the effective theory the second symmetry remains unbroken in both
 minima. 
The corresponding flavor current determined by the Wess--Zumino
term~(\ref{our}) is
\ba
\label{ourfl}
j^{\mu}&=&{1\over 4\pi^2\Lambda^{4N}}\det\! M\cdot
\epsilon^{\mu\nu\lambda\rho} \d_{\nu}B\;\;\!\!\d_{\lambda}\tilde{B}\Tr\l
T^f\d_{\rho}MM^{-1}\r\nonumber\\&&+h.c.  \ea
 Therefore, this current is zero in
both stationary points. Nevertheless, in the minimum with unbroken
baryon symmetry (and in the supersymmetric case as well) the charge
matching condition corresponding to the additional symmetry is
satisfied.  The point is that fermionic components of mesons ${\psi_{\sm
M}}_{\tilde{i}}^j$ are charged under this symmetry provided that
one of
the indices $(\tilde{i},j)$ is less than 
or equal to $N_0$, while another
is larger than $N_0$.  In the first candidate vacuum mesons obtain
non-zero vacuum expectation values (\ref{vevs}) which generate
spatially dependent masses for  ${\psi_{\sm
M}}_{\tilde{i}}^j$,
$${\cal A}{\d^2\det\! M\over\d M^j_{\tilde i}
\d M^k_{\tilde l}}{\psi_{\sm
M}}_{\tilde{i}}^j{\psi_{\sm
M}}_{\tilde{l}}^k
$$ 
through the term ${\cal A}\det\! M$ in the effective superpotential. Then
the calculation of the induced charge proceeds in the same way as in
the microscopic theory and one can straightforwardly check that the
resulting charge is again $N_c N[\tilde{U}]$. 

On the contrary, in the
minimum with broken baryon symmetry one has
$M^a_{\tilde{b}}({\bf x})=0$ and
the relevant fermions remain massless. Consequently, charge matching
conditions are not satisfied in this minimum, that strongly suggests
that the baryon symmetry is unbroken in the global 
vacuum~\footnote{It is worth noting,
however, that in the whole range ${\mu_B^2\over \mu_{M}^2}<{N\over2}$
there is a {\it local} vacuum in which matching conditions are
satisfied. For ${\mu_B^2\over \mu_{M}^2}>1$ this is the vacuum with
unbroken baryon symmetry and for ${\mu_B^2\over \mu_{M}^2}<1$ there
appears an
additional local minimum where both baryon and flavor symmetries are
broken.} and that
${\mu_B^2\over \mu_{M}^2}>{N\over 2}$. It should be noted, however,
that $\mu_B^2$ may be negative (it is the case for larger
$N_f$\cite{hamed}) and then higher order terms in the effective
potential~(\ref{softlow}) become essential. In any case, the above
consideration indicates that $SU(N_f)_L\times SU(N_f)_R$ flavor group
should be broken in the true vacuum of the theory.

\section{Conclusions}
\label{Conclusions}
We have considered matching of  induced charges in $SU(N)$ SQCD with
quantum modified moduli space. We have found that matching conditions
are satisfied provided an additional term similar to the Wess--Zumino
term in the non-supersymmetric QCD is added. Our calculation has shown 
that,
contrary to the QCD case, the third homotopy group $\pi_3({\cal Q})$
of the vacuum space is trivial. Consequently, no topological solitons
exist in SQCD.  Another consequence of this fact is that in the
supersymmetric case, the 
baryonic current carried by spatially inhomogeneous meson
fields is not  topological  and can be obtained
by making use of
the conventional Noether procedure. The second homotopy group
$\pi_2({\cal Q})$ was found to be also trivial,
so
there are no
global vortices in this theory either.

We have studied the case of softly broken supersymmetry as
well. In similarity  to the models considered in
Refs.~\cite{rubakov,weandru}, induced charge matching conditions
provide non-trivial information about the low-energy effective
theory in this case. Namely, they strongly disfavor the existence of
the global completely flavor symmetric vacuum. As a consequence, we
have obtained the lower bound on the ratio of the soft baryonic and mesonic
masses.

\section{Acknowledgments}
The authors are grateful to M.V.~Feigin, S.A.~Loktev and V.A.~Rubakov for
useful discussions.  This work is supported in part by Russian
Foundation for Basic Research grant 99-02-18410a, by the INTAS grant
96-0457 within the research program of the International Center for
Fundamental Physics in Moscow, by the Russian Academy of Sciences, 
JRP grant 37 and by ISSEP fellowships.
\section*{Appendix A. Topology of the quantum moduli space}
In this Appendix we present a proof of the homotopic equivalence given
by Eq.~(\ref{susp}). Let us first recall that the suspension $\Sigma
({\cal X})$ of the manifold $\cal X$ is the cylinder ${\cal X}\times
[0,1]$ where all points on the lower base ${\cal X}\times 0$ are
identified and all  points on the upper base ${\cal X}\times 1$ are
identified as well. As an example, the suspension of the
$d$-dimensional sphere $S^d$ is the $(d+1)$-dimensional sphere
$S^{(d+1)}$. The latter observation is the basis of the generalized
Freidental
theorem (``theorem about
suspension''),  which we refer to in the text.

In order to prove the homotopic equivalence~(\ref{susp}),
let us consider new variables $B_1$ and $B_2$ instead of $B,\tilde{B}$ such
that the quantum moduli space is determined by the following equation,
\[
\det\! M=\Lambda^{2N}-B_1^2-B_2^2.
\] 
Let us first consider  the manifold ${\cal Q}_1$ defined  by
the constraint
\be
\label{one}
\det\! M=\Lambda^{2N}-B_1^2, 
\ee 
and show that ${\cal Q}_1$ is homotopically
equivalent to $\Sigma (SL(N,\mathbb C))$. Indeed, the surface defined by 
Eq.~(\ref{one}) at a fixed value of $B_1\neq\pm\Lambda^N$ 
is topologically equivalent
to $SL(N,\mathbb C)$. At $B_1=\pm\Lambda^N$,
Eq.~(\ref{one}) defines the surfaces homotopically equivalent to a point.
Consequently, Eq.~(\ref{one})
indeed defines the manifold homotopically equivalent to $\Sigma
(SL(N,\mathbb C))$ if one restricts $B_1$ to belong to the interval
$[-\Lambda^N,\Lambda^N]$. Finally, let us construct
a deformation retract of ${\cal Q}_1$ onto its part determined by
$B_1\in[-\Lambda^N,\Lambda^N]$.  
\begin{figure}[tb]
\begin{center}
\psfig{file=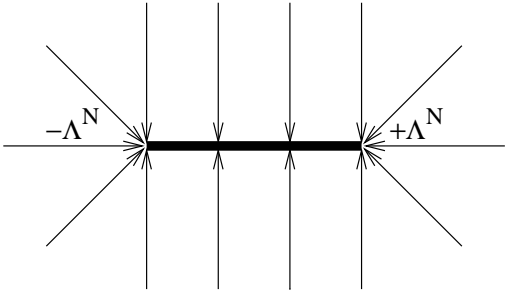}
\end{center}
\caption{{\protect\small Trajectories of the deformation
${\cal Q}_1$ $\to\Sigma (SL(N,\mathbb C))$ in the complex plane $B_1$.} }
\label{move}
\end{figure}
Let us take a point $(M^a_{\tilde{b}}(0),B_1(0))\in{\cal Q}_1$ with $B_1(0)\neq
\pm\Lambda^N$ . Let the coordinate $B_1$ move in the complex plane
as shown in Fig.~\ref{move} and define the matrix
$M^a_{\tilde{b}}(t)$ at the moment $t$ as follows,
\[
M(t)=TM(0)\;,
\]
where 
\[
T=\left(
\begin{array}{cc}
{\Lambda^{2N}-B_1(t)^2\over \Lambda^{2N}-B_1(0)^2} & 0 \\
0 &  \mathbf 1 \\
\end{array}\right)\;
\]
and $\mathbf 1$ is the $(N-1)\times (N-1)$ unit matrix.
The existence of such a deformation implies that ${\cal Q}_1\sim\Sigma
(SL(N,\mathbb C))$.

One can straightforwardly generalize these arguments and show
that ${\cal Q}\sim\Sigma ({\cal Q}_1)$. The last remark  to be
made for  proving the relation~(\ref{susp}) is that the $SU(N)$
group is homotopically equivalent to the $SL(N,\mathbb C)$ group.
\section*{Appendix B. Unambiguity of the Wess--Zumino term in 
 quantum theory}
$\Gamma_{SQCD}$ is unambiguous in  quantum theory iff it is 
equal to $2\pi n$, $n\in \mathbb Z$, when $\Sigma_5$ is the generator of
$\pi_5(\cal Q)$. The theorem about suspension implies the following
construction of this generator. Let us consider a
sphere $S^5$ with unit radius,
\[
x_1^2+x_2^2+x_3^2+x_4^2+x_5^2+x_6^2=1\; , \;\;\; x_i\in\mathbb R\;.
\]
The generator of $\pi_5(\cal Q)$ is the map 
$S^5\to\cal Q$ 
that in the notations of Appendix A can be written as follows
\[
B_1=\Lambda^Nx_5\;,
\]
\[
B_2=\Lambda^Nx_6\;,
\]
\[
M^a_{\tilde{b}}=\Lambda^2(1-x_5^2-x_6^2)^{1\over N}U^a_{\tilde{b}}
(x_1,x_2,x_3,x_4)\;,
\]
where $U^a_{\tilde{b}}(x_1,x_2,x_3,x_4)$ defines a generator of
$\pi_3(SU(N))$. Taking into account the expression~(\ref{NU}) for the
topological number of the map $S^3\to SU(N)$ and recalling 
that $dB\land
d\tilde{B}=2idB_1\land dB_2$ one obtains 
\[
\Gamma_{SQCD}=4\int_{x_5^2+x_6^2\le1}dx_5dx_6(1-x_5^2-x_6^2)=2\pi
\]
when $\Sigma_5$ is the generator of $\pi_5(\cal Q)$. Consequently, the
Wess--Zumino term (\ref{our}) is indeed unambiguous in  quantum
theory.
\def\ijmp#1#2#3{{\it Int. Jour. Mod. Phys. }{\bf #1~}(19#2)~#3}
\def\pl#1#2#3{{\it Phys. Lett. }{\bf B#1~}(19#2)~#3}
\def\zp#1#2#3{{\it Z. Phys. }{\bf C#1~}(19#2)~#3} 
\def\prl#1#2#3{{\it Phys. Rev. Lett. }{\bf #1~}(19#2)~#3} 
\def\rmp#1#2#3{{\it Rev. Mod. Phys. }{\bf #1~}(19#2)~#3} 
\def\prep#1#2#3{{\it Phys. Rep.    }{\bf #1~}(19#2)~#3} 
\def\pr#1#2#3{{\it Phys. Rev. }{\bf    D#1~}(19#2)~#3} 
\def\np#1#2#3{{\it Nucl. Phys. }{\bf    B#1~}(19#2)~#3} 
\def\mpl#1#2#3{{\it Mod. Phys. Lett. }{\bf    #1~}(19#2)~#3} 
\def\arnps#1#2#3{{\it Annu. Rev. Nucl. Part. Sci.    }{\bf #1~}(19#2)~#3} 
\def\sjnp#1#2#3{{\it Sov. J. Nucl. Phys.    }{\bf #1~}(19#2)~#3} 
\def\jetp#1#2#3{{\it JETP Lett. }{\bf    #1~}(19#2)~#3} 
\def\app#1#2#3{{\it Acta Phys. Polon. }{\bf    #1~}(19#2)~#3} 
\def\rnc#1#2#3{{\it Riv. Nuovo Cim. }{\bf    #1~}(19#2)~#3} 
\def\ap#1#2#3{{\it Ann. Phys. }{\bf #1~}(19#2)~#3}
\def\ptp#1#2#3{{\it Prog. Theor. Phys. }{\bf #1~}(19#2)~#3}
\def\spu#1#2#3{{\it Sov. Phys. Usp.}{\bf #1~}(19#2)~#3}
\addcontentsline{toc}{section}{Литература}

\end{document}